Version 6

# Resilience reconsidered: Case histories from disease ecology

Rodrick Wallace, Ph.D.
The New York State Psychiatric Institute
Deborah N. Wallace, Ph.D.
Dept. of Sociomedical Sciences
Mailman School of Public Health, Columbia University *

October 30, 2003

## Abstract

We expand the current understanding of ecological resilience to include the nested hierarchy of cognitive submodules that particularly, if not uniquely, delineates human ecosystems. These modules, ranging from the immune system to the local social network, are embedded in a cultural milieu which, to take the perspective of the evolutionary anthropologist Robert Boyd, "is as much a part of human biology as the enamel on our teeth". We begin by extending recent treatment of cognitive process, and of embedding culture itself, as associated with characteristic information sources, to a certain class of ecosystems through a generalization of coarse-graining. In the spirit of the Large Deviations Program, we then import renormalization formalism via the Asymptotic Equipartition Theorem of information theory to obtain punctuated response to parameters of increasing habitat degradation. A Legendre transform of an appropriate joint information permits analysis away from critical points, and generates the expected resilient quasi-stability in a highly natural manner. We interweave the discussion with applications to the public health impacts of the massive deurbanization and deindustrialization presently afflicting the United States.

## Introduction

We are compelled to begin with an extended and somewhat disturbing discussion of context: public health is under siege in the USA. Nearly a quarter-century into the AIDS pandemic, and antiretroviral therapies to the contrary, a relentless toll of about 40,000 new human immunosuppressive virus (HIV) infections continues annually (Rosenberg and Biggar, 1998). In addition, HIV has become resurgent among groups in which it had previously been contained (Wolitski et al., 2001). Asthma has been rising for decades, particularly within urban minority neighborhoods (e.g. Carr et al., 1992; Wright et al., 1999). Tuberculosis became resurgent after 1978 in New York City, the conurbation which most dominates the nation's urban hierarchy, a fact having dire implications for containment of the next wave of emerging infections (e.g. D. Wallace, 1994, 2001; Wallace and Wallace, 1997; Wallace et al., 1995; 1999). Obesity, already described as 'epidemic', has been increasing across all population subgroups for twenty years, and shows no signs of abatement (e.g. Flegal et al., 2002). 'Health disparities' between affluent majority and minority populations persist or widen in spite of massive health care expenditures (e.g. Barnett et al., 2001; Casper et al., 2000). Redirection of scarce resources into 'bioterrorist' activities, according to informed opinion, threatens to further damage the nation's already frail public health enterprise.

Three case histories across this increasingly bleak landscape will be the focus of our analysis: New York's tuberculosis outbreak, and the national obesity and asthma 'epidemics', as indexed by the annual diabetes death rate across all populations and the asthma death rate among African-Americans.

With regard to New York City, as has been described at greater length elsewhere (e.g. R. Wallace and D. Wallace, 1997; D. Wallace and R. Wallace, 1998; Duryea, 1978), cutbacks in essential municipal services during the 1970's, part of an extra-constitutional 'planned shrinkage' program apparently aimed at both dispersing minority voting blocks and clearing land for commercial purposes, initiated a very large and destructive outbreak of contagious building fire and housing abandonment in high population density, poverty-stricken minority neighborhoods. While the 'South Bronx' became synonymous with the ensuing disaster, large parts of Central Harlem in Manhattan, and a vast band across the minority sections of Northern Brooklyn, were badly affected as well. By 1980 some 600,000 people were living in devastated zones of the city (McCord and Freeman, 1990), and about 1.2 million non-Hispanic whites had fled the city for the suburbs. After 1978, in the face of what amounted to wartime conditions, tuberculosis cases began to rise in New York City, by 1985 the disease was clearly resurgent, and by 1990 the city's outbreak had begun to affect outlying suburban populations (e.g. Wallace et al., 1995, 1997).

*Address correspondence to: R. Wallace, PISCS Inc., 549 W. 123 St., Suite 16F, New York, NY, 10027. Telephone (212) 865-4766, rdwall@ix.netcom.com. Affiliations are for identification only.

By 1980, adult male life expectancy in Central Harlem had declined to less than that of adult males in Bangladesh (McCord and Freeman, 1990).

Figure 1 displays, for the Bronx section of New York City, the change in percent of occupied housing units between 1970 and 1980. The data are shown by 'Health Area', standard aggregates of US Census Tracts by which morbidity and mortality statistics are reported in the city. A huge chunk of the South-Central section lost as much as 81 percent of its occupied housing in a short period.

The Bronx, with 1.4 million people in 1970, is one of the largest urban centers in the industrialized world. Such peacetime devastation is unprecedented.

Figure 2 is an annual composite index of building fire and seriousness in New York City for the period 1959-1990, constructed by principal component analysis of numbers of structural fires, 'serious' fires requiring five or more units for extinguishment, and 'extra alarm assignments' beyond the first responding set of units. The index has been shifted so that zero fires gives a zero value. This is an environmental index of the contagious urban decay which produced figure 1 (Wallace and Wallace, 1998).

Figure 3 plots the number of reported tuberculosis cases in New York City against the *integral* of figure 2 from 1959. The argument is that, once a community is destroyed by contagious urban decay, the social, economic, and political capital contained in it are permanently dispersed, at least on the time scale of this analysis.

Two distinctly different systems are evident on this graph: a declining relation from 1959 through 1978, and a sharply rising one thereafter.

To anticipate the argument slightly, this sudden switch represents a kind of phase transition between two domains of relation in the tuberculosis ecosystem, with accumulated community damage constituting the driving parameter of a phase change having considerable hysteresis.

Figure 4 shows the counties of the Northeastern US which lost more than 1000 manufacturing jobs between 1972 and 1987, i.e. the deindustrialized 'rustbelt' which Ullmann (1988), Melman (1971) and others claim is a consequence of the massive diversion of scientific and engineering resources from civilian to military enterprise during the Cold War. They argue that a healthy US manufacturing economy required at least a 3 percent annual improvement in productivity to maintain itself against foreign competition, a rate of growth fueled by our considerable technical resources through about 1965. Thereafter, increasing consumption of scientific and engineering personnel by the military and aerospace enterprises of the Cold War, and the consequent shift in technological emphasis away from the needs of the civilian economy, made this rate of improvement impossible, and rapid US deindustrialization ensued. Wallace et al. (1999) examine the consequences of this collapse for the US AIDS epidemic.

Figure 4, like figure 1, represents the permanent dispersal of social, economic, and political capital within the worst affected counties, particularly as deindustrialization apparently contributed to patterns of deurbanization analogous to, if less acute than, that of figure 1 in many US cities, e.g. Pittsburgh, Dayton, Cleveland, Toledo, Philadelphia, Detroit, Gary, Akron, St. Louis, parts of Chicago, and so on.

Pappas, (1989) describes the general circumstances:

> "By 1982 mass unemployment had reemerged as a major social issue [in the USA]. Unemployment rose to its highest level since before World War II, and an estimated 12 million people were out of work – 10.8 percent of the labor force in the nation. It was not, however, a really new phenomenon. After 1968 a pattern was established in which each recession was followed by higher levels of unemployment during recovery. During the depth of the 1975 recession, national unemployment rose to 9.2 percent. In 1983, when a recovery was proclaimed, unemployment remained at 9.5 percent annually.
>
> Certain sectors of the work force have been more heavily affected than others. There was a 16.9 percent jobless rate among blue-collar workers in April, 1982... Unemployment and underemployment have become major problems for the working class. While monthly unemployment figures rise and fall, these underlying problems have persisted over a long period. Mild recoveries merely distract out attention from them."

Massey (1990) explores the particularly acute effect of this phenomenon on minorities:

> "The decline of manufacturing, the suburbanization of blue-collar employment, and the rise of the service sector eliminated many well-paying jobs for unskilled minorities and reduced the pool of marriageable men, thereby undermining the strength of the family, increasing the rate of poverty, and isolating many inner-city residents from accessible, middle-class occupations."

Figure 5 shows the number of manufacturing jobs in the US between 1980 and 2001, and figure 6 relates the *integral of manufacturing job loss after 1980* to the national diabetes death rate per 100,000 between 1980 and 1998. Diabetes is, of course, a quintessential index of obesity, a problem linked to a broad spectrum of chronic diseases, not only diabetes, but including coronary heart disease, asthma, and certain forms of cancer.

Two regions are evident in figure 6: a rough plateau between 1980 and 1988, followed by a sudden jump, and a sharp, nearly linear, rise thereafter.

Figure 7 is an analogous graph for the annual asthma death rate among African-Americans, plotted against the same jobs deficit as figure 6 over the same period. The graph shows a relentless 'inverted-J' transition from a relatively low to a much higher endemic rate. The change has taken place in the face of significant attempts at medical intervention, although these have primarily been exhortations of the sick to 'take responsibility and manage your disease', the processes of figures 1 and 4 notwithstanding. As the classic paper of Carr et al. (1992) shows, most of the asthma increase has been focused in minority urban communities which, we now know, have particularly suffered the synergism of those processes.

We will argue here that the changes in these two chronic diseases are an ecosystem phase transition recognizably similar to that shown in figure 3, with the dispersal of social, economic, and political capital indexed by the integral of annual manufacturing job loss representing a kind of human habitat desertification serving as the driving parameter of that transition.

Ecosystem theorists recognize several different kinds of 'resilience' (e.g. Gunderson, 2000). The first, which they call 'engineering resilience', since it is particularly characteristic of machines and man-machine interactions, involves the rate at which a disturbed system returns to a presumed single, stable, equilibrium condition, following perturbation. From that limited perspective, a resilient system is one which quickly returns to its one stable state.

Not many biological phenomena – including those of human physiology and psychology – are actually resilient in this simplistic sense.

Holling's (1973) particular contribution was to recognize that sudden transitions between different, at best quasi-stable, domains of relation among ecosystem variates were possible, i.e. that more than one 'stable' state was possible for real ecosystems. Gunderson (2000) puts the matter as follows:

> "One key distinction between these two types of resilience lies in assumptions regarding the existence of multiple [quasi-]stable states. If it is assumed that only one stable state exists or can be designed to exist, then the only possible definition and measures for resilience are near equilibrium ones – such as characteristic return time... The concept of ecological resilience presumes the existence of multiple stability domains and the tolerance of the system to perturbations that facilitate transitions among stable states. Hence, ecological resilience refers to the width or limit of a stability domain and is defined by the magnitude of disturbance that a system can absorb before it changes stable states... The presence of multiple [quasi-]stable states and transitions among them [has] been [empirically] described in a [large] range of ecological systems..."

Figure 3 shows two patterns of relation between accumulated community damage and tuberculosis: monotonic decline before 1978, and monotonic increase thereafter. Figure 6 shows stability of diabetes death rate with accumulated deindustrialization between 1980 and 1988, and a jump and linear rise thereafter, while figure 7 appears to be an inverted-J transition to a raised endemic level.

An often presumed difference between 'natural' and human-dominated ecosystems is, however, the particular role of both individual and collective cognitive action: human ecosystems are not simply reflex-driven, but can oppose reasoned, organized, responses to perturbation. A city facing a fire/housing abandonment epidemic or a nation undergoing deindustrialization are not simply ponds downhill from a leaking sewage treatment plant which are subject to sudden eutrophication.

This paper presents a treatment of ecosystem resilience which can, perhaps most crucially, be extended in a highly natural manner to include a large class of explicitly cognitive individual and collective phenomena, and indeed, to certain cognitive submodules of human biology. The expanded analysis indeed recovers most of classical ecosystem resilience theory, suggesting that the appropriate integrals of figures 2 and 5 do, in fact, parametize permanent, or at least highly persistent, change in essential institutions and in the relations of individuals and their health to those institutions, as indexed by figures 3, 6, and 7.

Some comment on methodology is appropriate.

Jimenez-Montano (1989) describes the language metaphor of theoretical biology, which we adopt here, as follows:

> "In his epilogue to the fourth volume of papers issuing from the IUBS Symposia at Villa Serbellion, inspired by papers of Pattee (1972) and Thom (1972), among others, Waddington concluded that...[in] situations which arise when there is mutual interaction between the complexity-out-of-simplicity (self-assembly), and simplicity-out-of-complexity (self-organization) processes are...to be discussed most profoundly at the present time with the help of the analogy of language...
>
> ...Waddington [thus] suggested that it is a *language* that may become a paradigm for [a Theory of General Biology], but a language in which basic sentences are programs, not statements..."

Jimenez-Montano goes on to summarize the elegant, if rather limited, applications of formal language theory to molecular genetics, quoting Pattee (1972) to the effect that a molecule becomes a message only in the context of larger constraints which he has called a 'language.'

Perhaps the essential defining constraint of any language is that not all possible sequences of symbols are meaningful, that is, systems of symbols must be arranged according to grammar and syntax in the context of even higher levels of structure, a matter to which we will repeatedly return.

The full implications of the Shannon-McMillan Theorem (SMT), otherwise known as the Asymptotic Equipartition Theorem, for Waddington's program seem to have been largely overlooked. A development based on that theorem is useful, in particular, for examining the ways in which vastly different 'languages' can interact to create complicated structures. The central problem then becomes the characterization of ecosystems and the cognitive processes of human biology and sociology as such (generalized) languages.

Our approach, based on robust application of the SMT, has a flavor recognizably similar to that surrounding the role of the Central Limit Theorem (CLT) in parametric statistics. Regardless of the probability distribution of a particular stochastic variate, the CLT ensures that long sums of independent realizations of that variate will follow a Normal distribution. Analogous constraints exist on the behavior of the information sources – both independent and interacting – that we will infer characterize human and natural ecosystems, and these are described by the SMT. Imposition of phase transition formalism from statistical physics onto the SMT, in the spirit of the Large Deviations Program of applied probability, permits concise and unified description of evolutionary and cognitive 'learning plateaus' which, in the evolutionary case, are interpreted as evolutionary punctuation (e.g. Wallace, 2002a, b). This perspective provides a 'natural' means of exploring punctuated processes in the effects of progressive habitat disruption on individuals, institutions, and their interactions.

Punctuated biological processes are found across temporal scales. Evolutionary punctuation is a modern extension of Darwinian evolutionary theory that accounts for the relative stability of a species' fossil record between the time it first appears and its extinction (e.g. Eldredge, 1985; Gould, 2002). Species appear 'suddenly' on a geologic time scale, persist relatively unchanged for a fairly long time, and then disappear sudden, again on a geologic time scale. Evolutionary process is vastly speeded up in tumorigenesis, which nonetheless seems amenable to similar analysis (e.g. Wallace et al., 2003).

Our model, as in the relation of the CLT to parametric statistical inference, is almost independent of the detailed structure of the interacting information sources inevitably associated with both cognitive and ecosystem processes, important as such structure may be in other contexts. This finesses some of the profound ambiguities associated with 'dynamic systems theory' and 'deterministic chaos' treatments in which the existence of 'dynamic attractors' depends on very specific kinds of differential equation models akin to those used to describe test-tube population dynamics, chemical processes, or physical systems of weights-on-springs. Cognitive and natural ecosystem phenomena are neither well-stirred Erlenmeyer flasks of reacting agents, nor distorted, noisy, clocks, and the application of 'nonlinear dynamic systems theory' to cognition or ecology will, in all likelihood, ultimately be found to involve little more than hopeful metaphor. Indeed, as shown below, much of nonlinear dynamics can be subsumed within an information theory formalism through symbolic dynamics 'coarse-graining' discretization techniques (e.g. Beck and Schlogl, 1995).

While idiosyncratic approaches analogous to nonparametric models in statistical theory may be required in some cases, the relatively straightforward formalism we develop here, like its cousin of parametric statistics, may well robustly capture the essence of much relevant phenomena.

### Modeling resilience

**1. Coarse-graining a simple ecosystem model** We begin with a simplistic picture of an elementary predator/prey ecosystem which, nonetheless, provides a useful pedagogical starting point. Let $X$ represent the appropriately scaled number of predators, $Y$ the scaled number of prey, $t$ the time, and $\omega$ a parameter defining their interaction. The model assumes that the ecologically dominant relation is an interaction between predator and prey, so that

$$dX/dt = \omega Y$$

$$dY/dt = -\omega X$$

Thus the predator populations grows proportionately to the prey population, and the prey declines proportionately to the predator population.

After differentiating the first and using the second equation, we obtain the differential equation

$$d^2X/dt^2 + \omega^2 X = 0$$

having the solution

$$X(t) = sin(\omega t); Y(t) = cos(\omega t).$$

with

$$X(t)^2 + Y(t)^2 = sin^2(\omega t) + cos^2(\omega t) \equiv 1.$$

Thus in the two dimensional 'phase space' defined by $X(t)$ and $Y(t)$, the system traces out an endless, circular trajectory in time, representing the out-of-phase sinusoidal oscillations of the predator and prey populations.

Divide the $X - Y$ 'phase space' into two components – the simplest 'coarse graining' – calling the halfplane to the left of the vertical $Y$-axis $A$ and that to the right $B$. This system, over units of the period $1/(2\pi\omega)$, traces out a stream of $A$'s and $B$'s having a very precise 'grammar' and 'syntax', i.e.

$$ABABABAB...$$

Many other such 'statements' might be conceivable, e.g.

$$AAAAA..., BBBBB..., AAABAAAB..., ABAABAAAB...,$$

and so on, but, of the obviously infinite number of possibilities, only one is actually observed, is 'grammatical', i.e. $ABABABAB...$.

More complex dynamical system models, incorporating diffusional drift around deterministic solutions, or even very elaborate systems of complicated stochastic differential equations, having various 'domains of attraction', i.e. different sets of grammars, can be described by analogous 'symbolic dynamics' (e.g. Beck and Schlogl, 1993, Ch. 3).

**2. Ecosystems as information sources** Rather than taking symbolic dynamics as a simplification of 'more exact' analytic or stochastic approaches, it proves useful, as it were, to throw out the Cheshire cat, but keep the cat's smile, generalizing symbolic dynamics to a more comprehensive information dynamics: Ecosystems may not have identifiable sets of dynamic equations like noisy, nonlinear clocks, but, under appropriate coarse-graining, they may still have recognizable sets of grammar and syntax over the long-term.

Examples abound: the turn-of-the seasons in a temperate climate, for many communities, looks remarkably the same year after year: the ice melts, the migrating birds return, the trees bud, the grass grows, plants and animals reproduce, high summer arrives, the foliage turns, the birds leave, frost, snow, the rivers freeze, and so on.

Suppose it is indeed possible to empirically characterize an ecosystem at a given time $t$ by observations of both habitat parameters such as temperature and rainfall, and numbers of various plant and animal species.

Traditionally, one can then calculate a cross-sectional species diversity index at time $t$ using an 'information' or 'entropy' metric of the form

$$H = -\sum_{j=1}^{M} (n_j/N) \log[(n_j/N)],$$

$$N \equiv \sum_{j=1}^{M} n_j$$

(1)

where $n_j$ is the number of observed individuals of species $j$ and $N$ is the total number of individuals of all species observed (e.g. Pielou, 1977; Ricotta, 2003; Fath et al., 2003).

This is not the approach taken here. Quite the contrary, in fact. Suppose it is possible to 'coarse grain' the ecosystem at time $t$ according to some appropriate partition of the 'phase space' in which each division $A_j$ represent a particular range of numbers of each possible species in the ecosystem, along with associated parameters such as temperature, rainfall, and the like. What is of particular interest to our development is not cross sectional structure, but rather longitudinal 'paths', i.e. ecosystem 'statements' of the form

$$x(n) = A_0, A_1, ..., A_n$$

defined in terms of some 'natural' time unit of the system, i.e. $n$ corresponds to an again appropriate characteristic time unit $T$, so that $t = T, 2T, ..., nT$.

To reiterate, unlike the traditional use of 'information theory' in ecology, our interest is in the *serial correlations along paths*, and not at all in the cross-sectional 'entropy' calculated for of a single element of a path.

Let $N(n)$ be the number of possible paths of length $n$ which are consistent with the underlying grammar and syntax of the 'appropriately coarsegrained' ecosystem, e.g. spring leads to summer, autumn, winter, back to spring, etc. but never something of the form spring to autumn to summer to winter in a temperate ecosystem.

The fundamental assumptions are that – for this chosen coarse-graining – $N(n)$, the number of possible grammatical paths, is much smaller than the total number of paths possible, and that, in the limit of (relatively) large $n$,

$$H = \lim_{n \to \infty} \frac{\log[N(n)]}{n}$$

(2)

both exists and is independent of path.

This is a critical foundation to, and limitation on, the modeling strategy and its range of strict applicability, but is, in a sense, fairly general since it is *independent of the details of the serial correlations along a path*.

Again, these conditions are the essence of the parallel with parametric statistics. Systems for which the assumptions are not true will require special 'nonparametric' approaches. We are inclined to believe, however, that, as for parametric statistical inference, the methodology will prove robust in that many systems will 'sufficiently' fulfill the essential criteria.

This being said, some further comment does seem necessary. Not all possible ecosystem coarse-grainings are likely to work, and different such divisions, even when appropriate, might well lead to different 'languages' for the ecosystem of interest. The example of Markov models is relevant. The essential Markov assumption is that the probability of a transition from one state at time $T$ to another at time $T + \Delta T$ depends only on the state at $T$, and not at all on the history by which that state was reached. If changes within the interval of length $\Delta T$ are 'plastic', or 'path dependent', then attempts to model the system as a Markov process *within* the 'natural' interval $\Delta T$ will fail, even though the model works quite well for phenomena separated by 'natural' intervals.

Thus empirical identification of relevant coarse-grainings for which our theory will work is clearly not trivial, and may, in fact, constitute the scientific core of the matter.

This is not, however, a new difficulty in ecosystem theory. Holling (1992), for example, explores the linkage of ecosystems across scales, finding that mesoscale structures – what might correspond to the neighborhood in a human community – are 'ecological keystones' in space, time, and population, which drive process and pattern at both smaller and larger scales and levels of organization. Levin (1989) writes

> "...[T]here is no single correct scale of observation: the insights one achieves from any investigation are contingent on the choice of scales. Pattern is neither a property of the system alone nor of the observer, but of an interaction between them... pattern exists at all levels and at all scales, and recognition of this multiplicity of scales is fundamental to describing and understanding ecosystems... there can be no 'correct' level of aggregation... We must recognize explicitly the multiplicity of scales within ecosystems, and develop a perspective that looks across scales and that builds on a multiplicity of models rather than seeking the single 'correct' one."

Given an appropriately chosen coarse-graining, whose selection in many cases will be the difficult and central trick of scientific art, suppose it possible to define joint and conditional probabilities for different ecosystem paths, having the form

$$P(A_0, A_1, ..., A_n), P(A_n|A_0, ..., A_{n-1})$$

(3)

such that appropriate joint and conditional Shannon uncertainties can be defined on them. For paths of length two these would be of the form

$$H(X_1, X_2) \equiv -\sum_j \sum_k P(A_j, A_k) \log[P(A_j, A_k)]$$

$$H(X_1|X_2) \equiv -\sum_j \sum_k P(A_j, A_k) \log[P(A_j|A_k)],$$

where the $X_j$ represent the stochastic processes generating the respective paths of interest.

The essential content of the Shannon-McMillan Theorem is that, for a large class of systems characterized as 'information sources', a kind of law-of-large numbers exists in the limit of very long paths, so that

$$H[X] = \lim_{n \to \infty} \frac{\log[N(n)]}{n} =$$

$$\lim_{n \to \infty} H(X_n|X_0, ..., X_{n-1}) =$$

$$\lim_{n \to \infty} \frac{H(X_0, X_1, ..., X_n)}{n+1}.$$

(4)

Taking the definitions of Shannon uncertainties as above, and arguing backwards from the latter two equations (e.g. Khinchine, 1957), it is indeed possible to recover the first, and divide the set of all possible temporal paths of our ecosystem into two subsets, one very small, containing the grammatically correct, and hence highly probable paths, which we will call 'meaningful', and a much larger set of vanishingly low probability.

Basic material on information theory can be found in any number of texts, e.g. Ash (1990), Khinchine (1957), Cover and Thomas (1991).

The next task is to show how the cognitive processes which so distinguish individual and collective human activity, can be fitted into a similar context, i.e. characterized as information sources.

**3. Cognition as an information source** Atlan and Cohen (1998) argue that the essence of cognition is comparison of a perceived external signal with an internal, learned picture of the world, and then, upon that comparison, the choice of one response from a much larger repertoire of possible responses.

Following the approach of Wallace (2000, 2002a), it is possible to make a very general model of this process as an information source.

Cognitive pattern recognition-and-selected response, as conceived here, proceeds by convoluting an incoming external 'sensory' signal with an internal 'ongoing activity' – the 'learned picture of the world' – and, at some point, triggering an appropriate action based on a decision that the pattern of sensory activity requires a response. It is not necessary to specify how the pattern recognition system is 'trained', and hence possible to adopt a weak model, regardless of learning paradigm, which can itself be more formally described by the Rate Distortion Theorem. Fulfilling Atlan and Cohen's (1998) criterion of meaning-from-response, we define a language's contextual meaning entirely in terms of system output.

The model, an extension of that presented in Wallace (2000), is as follows.

A pattern of 'sensory' input, say an ordered sequence $y_0, y_1, ...$, is mixed in a systematic way with internal 'ongoing' activity, the sequence $w_0, w_1, ...$, to create a path of composite signals $x = a_0, a_1, ..., a_n, ...$, where $a_j = f(y_j, w_j)$ for a function $f$. An explicit example will be given below. This path is then fed into a highly nonlinear 'decision oscillator' which generates an output $h(x)$ that is an element of one of two (presumably) disjoint sets $B_0$ and $B_1$. We take

$$B_0 \equiv b_0, ..., b_k,$$

$$B_1 \equiv b_{k+1}, ..., b_m.$$

Thus we permit a graded response, supposing that if

$$h(x) \in B_0$$

the pattern is not recognized, and if

$$h(x) \in B_1$$

the pattern is recognized and some action $b_j, k+1 \leq j \leq m$ takes place.

The principal focus of interest is those composite paths $x$ which trigger pattern recognition-and-response exactly once. That is, given a fixed initial state $a_0$, such that $h(a_0) \in B_0$, we examine all possible subsequent paths $x$ beginning with $a_0$ and leading exactly once to the event $h(x) \in B_1$. Thus $h(a_0, ..., a_j) \in B_0$ for all $j < m$, but $h(a_0, ..., a_m) \in B_1$.

For each positive integer $n$ let $N(n)$ be the number of paths of length $n$ which begin with some particular $a_0$ having $h(a_0) \in B_0$ and lead to the condition $h(x) \in B_1$. We shall call such paths 'meaningful' and assume $N(n)$ to be considerably less than the number of all possible paths of length $n$ – pattern recognition-and-response is comparatively rare. We further assume that the longitudinal finite limit

$$H \equiv \lim_{n \to \infty} \frac{\log[N(n)]}{n}$$

both exists and is independent of the path $x$. We will – not surprisingly – call such a cognitive process *ergodic*.

Note that disjoint partition of 'state space' may be possible according to sets of states which can be connected by 'meaningful' paths, leading to a 'natural' coset algebra of the system, a matter of some importance not pursued further here.

It is thus possible to define an ergodic information source **X** associated with stochastic variates $X_j$ having joint and conditional probabilities $P(a_0, ..., a_n)$ and $P(a_n|a_0, ..., a_{n-1})$ such that appropriate joint and conditional Shannon uncertainties may be defined which satisfy the relations of equation 4 above.

This information source is taken as *dual* to the ergodic cognitive process.

As stated, the Shannon-McMillan Theorem and its variants provide 'laws of large numbers' which permit definition of the Shannon uncertainties in terms of cross-sectional sums of the form

$$H = -\sum P_k \log[P_k],$$

where the $P_k$ constitute a probability distribution.

It is important to recognize that different 'languages' will be defined by different divisions of the total universe of possible responses into various pairs of sets $B_0$ and $B_1$, or by requiring more than one response in $B_1$ along a path. Like the use of different distortion measures in the Rate Distortion Theorem (e.g. Cover and Thomas, 1991), however, it seems obvious that the underlying dynamics will all be qualitatively similar.

Nonetheless, dividing the full set of possible responses into the sets $B_0$ and $B_1$ may itself require 'higher order' cognitive decisions by another module or modules, suggesting the necessity of 'choice' within a more or less broad set of possible 'languages of thought'. This would directly reflect the need to 'shift gears' according to the different challenges faced by the organism or organization. A critical problem then becomes the choice of a 'normal' zero-mode language among a very large set of possible languages representing the 'excited states' accessible to the system. This is a fundamental matter which mirrors, for isolated cognitive systems, the resilience arguments applicable to more conventional ecosystems, i.e. the possibility of more than one 'zero state' to a cognitive system. Identification of an 'excited' state as the zero mode becomes, then, a kind of generalized autoimmune disorder which can be triggered by linkage with external 'ecological' information sources of structured psychosocial stress, a matter we explore at length elsewhere (Wallace et al., 2003).

In sum, meaningful paths – creating an inherent grammar and syntax – have been defined entirely in terms of system response, as Atlan and Cohen (1998) propose.

This formalism can be applied to the stochastic neuron in a neural network: A series of inputs $y_i^j, i = 1, ...m$ from $m$ nearby neurons at time $j$ to the neuron of interest is convoluted with 'weights' $w_i^j, i = 1, ..., m$, using an inner product

$$a_j = \mathbf{y}^j \cdot \mathbf{w}^j \equiv \sum_{i=1}^{m} y_i^j w_i^j \quad (5)$$

in the context of a 'transfer function' $f(\mathbf{y}^j \cdot \mathbf{w}^j)$ such that the probability of the neuron firing and having a discrete output $z^j = 1$ is $P(z^j = 1) = f(\mathbf{y}^j \cdot \mathbf{w}^j)$.

Thus the probability that the neuron does not fire at time j is just $1 - P$. In the usual terminology the $m$ values $y_i^j$ constitute the 'sensory activity' and the $m$ weights $w_i^j$ the 'ongoing activity' at time $j$, with $a_j = \mathbf{y}^j \cdot \mathbf{w}^j$ and the path $x \equiv a_0, a_1, ..., a_n, ....$ A more elaborate example is given in Wallace (2002a).

A little work leads to a standard neural network model in which the network is trained by appropriately varying $\mathbf{w}$ through least squares or other error minimization feedback. This can be shown to replicate rate distortion arguments, as we can use the error definition to define a distortion function which measures the difference between the training pattern $y$ and the network output $\hat{y}$ as a function, for example, of the inverse number of training cycles, $K$. As we will discuss in another context, 'learning plateau' behavior emerges naturally as a phase transition in the mutual information $I(Y, \hat{Y})$ driven by the parameter $K$.

This leads eventually to parametization of the information source uncertainty of the dual information source to a cognitive pattern recognition-and-response with respect to one or more variates, writing, e.g. $H[\mathbf{K}]$, where $\mathbf{K} \equiv (K_1, ..., K_s)$ represents a vector in a parameter space. Let the vector $\mathbf{K}$ follow some path in time, i.e. trace out a generalized line or surface $\mathbf{K}(t)$. We will, following the argument of Wallace (2002b), assume that the probabilities defining $H$, for the most part, closely track changes in $\mathbf{K}(t)$, so that along a particular 'piece' of a path in parameter space the information source remains as close to memoryless and ergodic as is needed for the mathematics to work. Between pieces we will, below, impose phase transition characterized by a renormalization symmetry, in the sense of Wilson (1971). See Binney, et al. (1986) and Wallace et al. (2003) for a more complete discussion of the formal mathematics.

Such an information source can be termed 'adiabatically piecewise memoryless ergodic' (APME). To reiterate, the ergodic nature of the information sources is a generalization of the 'law of large numbers' and implies that the long-time averages we will need to calculate can, in fact, be closely approximated by averages across the probability spaces of those sources. This is no small matter.

The reader may have noticed parallels with Fodor's speculations on the 'language of thought' (e.g. Fodor, 1975, 1981, 1987, 1990, 1994, 1998, 1999) in which he proposes a Chomskian 'linguistically complete' background 'natural' language as the basis of mental function in humans. This work, in contrast, takes a 'weak' asymptotic approach which does not require linguistic completeness, but only that very long series of outputs may be characterized as the product of a 'dual' information source which is adiabatically, piecewise, memoryless ergodic. This allows importation of phase transition and other methods from statistical physics, done below, which provide a natural approach to the interaction of cognitive submodules or the incorporation of individual and collective cognitive phenomena into a more general ecosystem perspective, as we do here.

## 4. Some cognitive modules of human biology

It is possible to cast a broad spectrum of human biological, psychological, and social phenomena in a cognitive context.

**Immune function** Atlan and Cohen (1998) have proposed an information-theoretic cognitive model of immune function and process, a paradigm incorporating cognitive pattern recognition-and-response behaviors analogous to those of the central nervous system. This work follows in a very long tradition of speculation on the cognitive properties of the immune system (e.g. Tauber, 1998; Podolsky and Tauber, 1998; Grossman, 1989, 1992, 1993a, b, 2000).

From the Atlan/Cohen perspective, the meaning of an antigen can be reduced to the type of response the antigen generates. That is, the meaning of an antigen is functionally defined by the response of the immune system. The meaning of an antigen to the system is discernible in the type of immune response produced, not merely whether or not the antigen is perceived by the receptor repertoire. Because the meaning is defined by the type of response there is indeed a response repertoire and not only a receptor repertoire.

To account for immune interpretation Cohen (1992, 2000) has reformulated the cognitive paradigm for the immune system. The immune system can respond to a given antigen in various ways, it has 'options.' Thus the particular response we observe is the outcome of internal processes of weighing and integrating information about the antigen.

In contrast to Burnet's view of the immune response as a simple reflex, it is seen to exercise cognition by the interpolation of a level of information processing between the antigen stimulus and the immune response. A cognitive immune system organizes the information borne by the antigen stimulus within a given context and creates a format suitable for internal processing; the antigen and its context are transcribed internally into the 'chemical language' of the immune system.

The cognitive paradigm suggests a language metaphor to describe immune communication by a string of chemical signals. This metaphor is apt because the human and immune languages can be seen to manifest several similarities such as syntax and abstraction. Syntax, for example, enhances both linguistic and immune meaning.

Although individual words and even letters can have their own meanings, an unconnected subject or an unconnected predicate will tend to mean less than does the sentence generated by their connection.

The immune system creates a 'language' by linking two ontogenetically different classes of molecules in a syntactical fashion. One class of molecules are the T and B cell receptors for antigens. These molecules are not inherited, but are somatically generated in each individual. The other class of molecules responsible for internal information processing is encoded in the individual's germline.

Meaning, the chosen type of immune response, is the outcome of the concrete connection between the antigen subject and the germline predicate signals.

The transcription of the antigens into processed peptides embedded in a context of germline ancillary signals constitutes the functional 'language' of the immune system. Despite the logic of clonal selection, the immune system does not respond to antigens as they are, but to abstractions of antigens-in-context.

**Tumor control** The next cognitive submodule after the immune system appears to be a tumor control mechanism which may include 'immune surveillance', but clearly transcends it. Nunney (1999) has explored cancer occurrence as a function of animal size, suggesting that in larger animals, whose lifespan grows as about the 4/10 power of their cell count, prevention of cancer in rapidly proliferating tissues becomes more difficult in proportion to size. Cancer control requires the development of additional mechanisms and systems to address tumorigenesis as body size increases – a synergistic effect of cell number and organism longevity. Nunney concludes

> "This pattern may represent a real barrier to the evolution of large, long-lived animals and predicts that those that do evolve ... have recruited additional controls [over those of smaller animals] to prevent cancer."

In particular, different tissues may have evolved markedly different tumor control strategies. All of these, however, are likely to be energetically expensive, permeated with different complex signaling strategies, and subject to a multiplicity of reactions to signals, including those related to psychosocial stress. Forlenza and Baum (2000) explore the effects of stress on the full spectrum of tumor control, ranging from DNA damage and control, to apoptosis, immune surveillance, and mutation rate. Elsewhere (R. Wallace et al., 2003) we argue that this elaborate tumor control strategy, at least in large animals, must be at least as cognitive as the immune system itself, which is one of its components: some comparison must be made with an internal picture of a 'healthy' cell, and a choice made as to response: none, attempt DNA repair, trigger programmed cell death, engage in full-blown immune attack. This is, from the Atlan/Cohen perspective, the essence of cognition.

**The HPA axis** The hypothalamic-pituitary-adrenal (HPA) axis, the 'flight-or-fight' system, is clearly cognitive in the Atlan/Cohen sense. Upon recognition of a new perturbation in the surrounding environment, memory and brain or emotional cognition evaluate and choose from several possible responses: no action needed, flight, fight, helplessness (i.e. flight or fight needed, but not possible). Upon appropriate conditioning, the HPA axis is able to accelerate the decision process, much as the immune system has a more efficient response to second pathogenic challenge once the initial infection has become encoded in immune memory. Certainly 'hyperreactivity' in the context of post-traumatic stress disorder (PTSD) is a well known example. Chronic HPA axis activation is deeply implicated in visceral obesity leading to diabetes and heart disease (e.g. Bjorntorp, 2001).

**Blood pressure regulation** Rau and Elbert (2001) review much of the literature on blood pressure regulation, particularly the interaction between baroreceptor activation and central nervous function. We paraphrase something of their analysis. The essential point, of course, is that unregulated blood pressure would be quickly fatal in any animal with a circulatory system, a matter as physiologically fundamental as tumor control. Much work over the years has elucidated some of the mechanisms involved: increase in arterial blood pressure stimulates the arterial baroreceptors which in turn elicit the baroreceptor reflex, causing a reduction in cardiac output and in peripheral resistance, returning pressure to its original level. The reflex, however, is not actually this simple: it may be inhibited through peripheral processes, for example under conditions of high metabolic demand. In addition, higher brain structures modulate this reflex arc, for instance when threat is detected and fight or flight responses are being prepared. This suggests, then, that blood pressure control cannot be a simple reflex, but is, rather, a broad and actively cognitive modular system which compares a set of incoming signals with an internal reference configuration, and then chooses an appropriate physiological level of blood pressure from a large repertory of possible levels, i.e. a cognitive process in the Atlan/Cohen sense. The baroreceptors and the baroreceptor reflex are, from this perspective, only one set of a complex array of components making up a larger and more comprehensive cognitive blood pressure regulatory module.

**Emotion** Thayer and Lane (2000) summarize the case for what can be described as a cognitive emotional process. Emotions, in their view, are an integrative index of individual adjustment to changing environmental demands, an organismal response to an environmental event that allows rapid mobilization of multiple subsystems. Emotions are the moment-to-moment output of a continuous sequence of behavior, organized around biologically important functions. These 'lawful' sequences have been termed 'behavioral systems' by Timberlake (1994).

Emotions are self-regulatory responses that allow the efficient coordination of the organism for goal-directed behavior. Specific emotions imply specific eliciting stimuli, specific action tendencies including selective attention to relevant stimuli, and specific reinforcers. When the system works properly, it allows for flexible adaptation of the organism to changing environmental demands, so that an emotional response represents a *selection* of an appropriate response and the inhibition of other less appropriate responses from a more or less broad behavioral repertoire of possible responses. Such 'choice' leads directly to something closely analogous to the Atlan and Cohen language metaphor.

Damasio (1998) concludes that emotion is the most complex expression of homeostatic regulatory systems. The results of emotion serve the purpose of survival even in nonminded organisms, operating along dimensions of approach or aversion, of appetite or withdrawal. Emotions protect the subject organism by avoiding predators or scaring them away, or by leading the organism to food and sex. Emotions often operate as a basic mechanism for making decisions without the labors of reason, that is, without resorting to deliberated considerations of facts, options, outcomes, and rules of logic. In humans learning can pair emotion with facts which describe the premises of a situation, the option taken relative to solving the problems inherent in a situation, and perhaps most importantly, the outcomes of choosing a certain option, both immediately and in the future. The pairing of emotion and fact remains in memory in such a way that when the facts are considered in deliberate reasoning when a similar situation is revisited, the paired emotion or some aspect of it can be reactivated. The recall, according to Damasio, allows emotion to exert its pairwise qualification effect, either as a conscious signal or as nonconscious bias, or both, In both types of action the emotions and the machinery underlying them play an important regulatory role in the life of the organism. This higher order role for emotion is still related to the needs of survival, albeit less apparently.

Thayer and Friedman (2002) argue, from a dynamic systems perspective, that failure of what they term 'inhibitory processes' which, among other things, direct emotional responses to environmental signals, is an important aspect of psychological and other disorder. Sensitization and inhibition, they claim, 'sculpt' the behavior of an organism to meet changing environmental demands. When these inhibitory processes are dysfunctional – choice fails – pathology appears at numerous levels of system function, from the cellular to the cognitive.

Thayer and Lane (2000) also take a dynamic systems perspective on emotion and behavioral subsystems which, in the service of goal-directed behavior and in the context of a behavioral system, they see organized into coordinated assemblages that can be described by a small number of control parameters, like the factors of factor analysis, revealing the latent structure among a set of questionnaire items thereby reducing or mapping the higher dimensional item space into a lower dimensional factor space. In their view, emotions may represent preferred configurations in a larger 'state-space' of a possible behavioral repertoire of the organism. From their perspective, disorders of affect represent a condition in which the individual is unable to select the appropriate response, or to inhibit the inappropriate response, so that the response selection mechanism is somehow corrupted.

Gilbert (2001) suggests that a canonical form of such 'corruption' is the excitation of modes that, in other circumstances, represent 'normal' evolutionary adaptations.

**'Rational thought'** Although the Cartesian dichotomy between 'rational thought' and 'emotion' may be increasingly suspect, nonetheless humans, like many other animals, do indeed conduct individual rational cognitive decision-making as most of us would commonly understand it. Various forms of dementia involve, among other things, degradation in that ability.

**Sociocultural network** Humans are particularly noted for a hypersociality which inevitably enmeshes us all in group processes of decision, i.e. collective cognitive behavior within a social network, tinged by an embedding shared culture. For humans, culture is truly fundamental. Durham (1991) argues that genes and culture are two distinct but interacting systems of inheritance within human populations. Information of both kinds has influence, actual or potential, over behaviors, which creates a real and unambiguous symmetry between genes and phenotypes on the one hand, and culture and phenotypes, on the other. Genes and culture are best represented as two parallel lines or tracks of hereditary influence on phenotypes.

Much of hominid evolution seems to focus on an interweaving of genetic and cultural systems. Genes came to encode for increasing hypersociality, learning, and language skills. The most successful populations displayed increasingly complex structures that better aided in buffering the local environment (e.g. Bonner, 1980).

Successful human populations seem to have a core of tool usage, sophisticated language, oral tradition, mythology, music, magic, medicine, and religion, and decision making skills focused on relatively small family/extended family social network groupings. More complex social structures are built on the periphery of this basic object (e.g. Richerson and Boyd, 1995). The human species' very identity may rest on its unique evolved capacities for social mediation and cultural transmission. These are particularly expressed through the cognitive decision making of small groups facing changing patterns of threat and opportunity, processes in which we are all embedded and all participate.

Next we explore how different information sources can interact, in particular, how deterioration of an embedding ecosystem can entrain the cognitive phenomena of human biology.

**5. Interacting information sources: punctuated crosstalk** Suppose that a cognitive process at the individual or group levels of organization can be represented by a sequence of 'states' in time, the 'path' $x \equiv x_0, x_1, ....$ Similarly, we assume an embedding ecosystem with which that process interacts can also be represented by a path $y \equiv y_0, y_1, ....$ These paths are both very highly structured and, within themselves, are serially correlated and can, in fact, be rep-resented by 'information sources' **X** and **Y**. We assume the cognitive process and embedding ecosystem interact, so that these sequences of states are not independent, but are jointly serially correlated. We can, then, define a path of sequential pairs as $z \equiv (x_0, y_0), (x_1, y_1), ....$

The essential content of the Joint Asymptotic Equipartition Theorem version of the Shannon-McMillan Theorem is that the set of joint paths $z$ can be partitioned into a relatively small set of high probability which is termed *jointly typical*, and a much larger set of vanishingly small probability. Further, according to the JAEPT, the *splitting criterion* between high and low probability sets of pairs is the mutual information

$$I(X,Y) = H(X) - H(X|Y) = H(X) + H(Y) - H(X,Y)$$

(6)

where $H(X), H(Y), H(X|Y)$ and $H(X,Y)$ are, respectively, the Shannon uncertainties of $X$ and $Y$, their conditional uncertainty, and their joint uncertainty. Again, see Cover and Thomas (1991) or Ash (1990) for mathematical details. As stated above, the Shannon-McMillan Theorem and its variants permit expression of the various uncertainties in terms of cross sectional sums of terms of the form $-P_k \log[P_k]$ where the $P_k$ are appropriate direct or conditional probabilities. Similar approaches to neural process have been recently adopted by Dimitrov and Miller (2001).

The high probability pairs of paths are, in this formulation, all equiprobable, and if $N(n)$ is the number of jointly typical pairs of length $n$, then, according to the Shannon-McMillan Theorem and its 'joint' variants,

$$I(X,Y) = \lim_{n \to \infty} \frac{\log[N(n)]}{n}.$$

(7)

Generalizing the earlier language-on-a-network models of Wallace and Wallace (1998, 1999), suppose there is a 'coupling parameter' $P$ representing the degree of linkage between the cognitive human subsystem of interest and the structured 'language' of the embedding ecosystem, and set $K = 1/P$, following the development of those earlier studies. Then we have

$$I[K] = \lim_{n \to \infty} \frac{\log[N(K,n)]}{n}.$$

The essential 'homology' between information theory and statistical mechanics lies in the similarity of this expression with the infinite volume limit of the free energy density. If $Z(K)$ is the statistical mechanics partition function derived from the system's Hamiltonian, then the free energy density is determined by the relation

$$F[K] = \lim_{V \to \infty} \frac{\log[Z(K)]}{V}.$$

(8)

$F$ is the free energy density, $V$ the system volume and $K = 1/T$, where $T$ is the system temperature.

Various authors argue at some length (e.g. Wallace, 2003; Rojdestvensky and Cottam, 2000; Feynman, 1996) that this is indeed a systematic mathematical homology which, it can be shown, permits importation of renormalization symmetry into information theory. Imposition of invariance under renormalization on the mutual information splitting criterion $I(X,Y)$ implies the existence of phase transitions analogous to learning plateaus or punctuated evolutionary equilibria in the relations between cognitive mechanism and the embedding ecosystem. An extensive mathematical treatment of these ideas is presented elsewhere (e.g. Wallace, 2000, 2002a,b; Wallace et al., 2003). Much of the uniqueness of the system under study will be expressed in the 'renormalization relations' associated with that punctuation. See Wallace et al. (2003) for fuller discussion.

Elaborate developments are possible. From a the more limited perspective of the Rate Distortion Theorem, a selective corollary of the Shannon-McMillan Theorem, we can view the onset of a punctuated interaction between the cognitive process and embedding ecosystem as the literal writing of distorted image of those systems upon each other, Lewontin's (2000) 'interpenetration':

Suppose that two (adiabatically, piecewise memoryless, ergodic) information sources **Y** and **B** begin to interact, to 'talk' to each other, i.e. to influence each other in some way so that it is possible, for example, to look at the output of **B** – strings $b$ – and infer something about the behavior of **Y** from it – strings $y$. We suppose it possible to define a retranslation from the B-language into the Y-language through a deterministic code book, and call $\hat{\mathbf{Y}}$ the translated information source, as mirrored by **B**.

Define some distortion measure comparing paths $y$ to paths $\hat{y}$, $d(y, \hat{y})$ (Cover and Thomas, 1991). We invoke the Rate Distortion Theorem's mutual information $I(Y, \hat{Y})$, which is the splitting criterion between high and low probability pairs of paths. Impose, now, a parametrization by an inverse coupling strength $K$, and a renormalization symmetry representing the global structure of the system coupling.

Extending the analyses, triplets of sequences, $Y_1, Y_2, Z$, for which one in particular, here $Z$, is the 'embedding context' affecting the other two, can also be divided by a splitting criterion into two sets, having high and low probabilities respectively. The probability of a particular triplet of sequences is then determined by the conditional probabilities

$$P(Y_1 = y^1, Y_2 = y^2, Z = z) = \Pi_{j=1}^n p(y_j^1|z_j)p(y_j^2|z_j)p(z_j). \quad (9)$$

That is, $Y_1$ and $Y_2$ are, in some measure, driven by their interaction with $Z$.

For large $n$ the number of triplet sequences in the high probability set will be determined by the relation (Cover and Thomas, 1992, p. 387)

$$N(n) \propto \exp[nI(Y_1; Y_2|Z)], \quad (10)$$

where splitting criterion is given by

$$I(Y_1; Y_2|Z) \equiv$$

$$H(Z) + H(Y_1|Z) + H(Y_2|Z) - H(Y_1, Y_2, Z).$$

It is then possible to examine mixed cognitive/adaptive phase transitions analogous to learning plateaus (Wallace, 2002b) in the splitting criterion $I(Y_1, Y_2|Z)$. We reiterate that these results are almost exactly parallel to the Eldredge/Gould model of evolutionary punctuated equilibrium (Eldredge, 1985; Gould, 2002).

The model is easily extended to any number of interacting information sources, $Y_1, Y_2, ..., Y_s$ conditional on an external context $Z$ in terms of a splitting criterion defined by

$$I(Y_1; ...; Y_s|Z) = H(Z) + \sum_{j=1}^{s} H(Y_j|Z) - H(Y_1, ..., Y_s, Z), \quad (11)$$

where the conditional Shannon uncertainties $H(Y_j|Z)$ are determined by the appropriate direct and conditional probabilities.

In general, then, it seems fruitful to characterize the mutual interpenetration of cognitive biopsychosocial and non-cognitive ecosystem structures of human community within the context a single, unifying, formal perspective summarized by a 'larger' information source, more precisely, invoking a mutual information between cognitive and ecosystem information sources.

The next step is to parametize the 'richness' of the interacting 'languages' by the degree of habitat degradation resulting from, for example, contagious urban decay or progressive deindustrialization, and examine likely patterns of dynamic change, which we will find to be both highly punctuated and constrained by significant irreversibility criteria: i.e. Humpty-Dumpty effects.

**6. Resilience quasi-thermodynamics** Our analysis suggests certain 'natural' ecosystems and more complex cognitive human communities (as well as individuals within communities and subsystems within individuals) may be characterized by 'languages' in a general sense, and has examined the way in which such information sources can interact in a punctuated manner to form composite entities. The next step is to explore phenomena of ecological resilience, in Holling's (1973) sense, for such composites, which will now be described in a simplified shorthand only as a single generalized source uncertainty $H$.

The essential assumption is that, as deindustrialization, deurbanization, or other processes of socioeconomic or political degradation proceed, the possible number of 'meaningful' paths for the composite system of interest becomes reduced. That is, the 'language' of the human ecosystem becomes less rich as social, economic, and political capital are dispersed by processes like those leading to figures 1 and 4.

Painful scenarios are not difficult to construct, but the detailed case studies of Pappas (1989), Fullilove (2004) and our own work (Wallace and Wallace, 1998) have already done a better job.

Suppose, then, that $H$ can be parametized by some index of human habitat degradation $K$, and suppose $H[K]$ a monotonically declining function of $K$. As Cover and Thomas (1991) or Ash (1990) describe, a decline in source uncertainty can be interpreted as a decline in underlying 'communication channel capacity', since $H \leq C$, where $C$ is that capacity. As social, economic, and political capital are dispersed by deurbanization or deindustrialization, the social channel erodes, and that, in turn, inevitably erodes the richness of the associated human ecosystem, both cognitive and other aspects: possible choices narrow, possible outcomes constrict.

If $H$ is the analog to free energy density in a physical system, $K$ is the analog of an inverse temperature, and the next

'natural' step is to apply a *Legendre transformation* to examine the behavior of the system *away* from phase transition, defining a rough analog to thermodynamics in a physical system. We will use the resulting formalism to explore a complicated, and sometimes irreversible, synergistic dynamic of habitat deterioration and ecosystem regime change.

The Legendre transform of a well-behaved function $f(Z_1, ..., Z_w)$, denoted $g$, is

$$g = f - \sum_{j=1}^{w} Z_j \partial f/\partial Z_j = f - \sum_{j=1}^{w} Z_j V_j,$$

$$V_j \equiv \partial f/\partial Z_j.$$

(12)

The relation is invertible everywhere provided $df/dZ$ is strictly monotonic. That is, if $d^2 f/dZ^2 > 0$ for all points, we can write

$$f = g - \sum_{k=1}^{w} V_k \partial g/\partial V_k.$$

The generalization when $f$ is not well-behaved is via a variational principle (e.g. Beck and Schlogl, 1993) rather than this tangent plane argument. We will suggest that, for structures strongly dominated by social or other biological interaction, the existence of points for which the monotonicity/convexity condition fails, e.g. points of inflection for $H$, may, in analogy to physical systems, imply the existence of systemic changes similar to phase transitions.

Consider a system for which $H$ monotonically declines with increasing $K$, an index of habitat deterioration. We define $S$, an entropy-analog which we term the 'disorder', as the Legendre transform of the source uncertainty $H$:

$$S = H - K dH/dK \equiv H - KU$$

(13)

where $U = dH/dK$ is like the 'internal energy' of a physical system. Since

$$dS/dK = dH/dK - U - K dU/dK = -K dU/dK,$$

then $dS/dU = -K$ and $dU \propto (1/K)dS$. This is like the classic thermodynamic relation $dQ = TdS$ for a physical system, if we take $1/K \to T$. Thus the disorder $S$ is indeed a generalized entropy, albeit a strange one.

Figure 8 shows a proposed smooth reverse S-shaped form for ecosystem source uncertainty as a function of $K$, an index of habitat degradation. As $K$ increases, the capacity of the ecosystem as a kind of communication channel declines, since possible choice narrows, and the number of possibly grammatical paths declines.

Parametize $H$ by $K$ and now invoke an interaction of the phase transition arguments from the sections above, *in terms of K*. Thus there will be a critical value, $K = K_0$, at which the ecosystem undergoes sudden transformation into a markedly different form of internal organization. See the appendix of Wallace et al. (2003) for arguments regarding the renormalization properties of such 'biological' phase transitions, as opposed to simple physical ones.

Of particular interest is in the dynamic behavior of the system as it approaches that ecological phase transformation.

Figure 8 shows the disorder $S = H - K dH/dK$ superimposed on $H$. Closed physical systems gravitate to positions at which an entropy measure, like $S$, is a maximum. We propose that (thermodynamically open) systems strongly dominated by social interaction and cognitive process are different from closed physical systems in that they will often act, insofar as they can, *to minimize the experience of disorder.*

For a physical system this kind of behavior can be described, to first order, by the Onsager relations which characterize the relation between the rate of change in driving system parameters $K_1, ..., K_m$ and the gradient of the entropy-measure in those parameters, i.e.

$$\sum_{k=1}^{m} R_{j,k} dK_k/dt = \partial S/\partial K_j$$

$$dK_j/dt = \sum_{k=1}^{m} L_{j,k} \partial S/\partial K_k$$

(14)

where the terms $\partial S/\partial K$ are the driving 'thermodynamic forces'. For physical systems the $L_{j,k}$ are positive. For social and biological systems, in contrast, it appears that there are generalized Onsager relations for which $L_{j,k}$ may be negative. That is, biological or social systems may be *repelled* by disorder (e.g. Wallace and Wallace, 1998, 1999).

Taking this perspective, a community with a low index of habitat deterioration, $K$, will begin at the left hand side of figure 8. As $K$ increases, it first raises the disorder $S$. Initial response of the ecosystem might be to attempt to compensate, and reorganize itself to lower the perceived disorder, driving the system back to the left, perhaps as a kind of limited homeostasis providing the often-observed ecological quasi-stability described by Gunderson (2000).

If, however, increases in $K$ are driven by overwhelming extrinsic factors of public policy or environmental change, then disorder rises to a maximum. Near that maximum, to the left, small increases in $K$ can cause a large increase in disorder. Just to the right of the maximum, however, small increases in $K$ can, paradoxically, cause large declines in the disorder $S$. Note that the maximum of $S$ represents a point of inflection for $H$ at which the monotonicity of $dH/dK$ fails. That is, $H$ is not necessarily a convex function, in sharp contrast to simple physical analogs.

At such a point we are led to suggest that a self-reinforcing dynamic or system of dynamics may emerge under which the larger community literally falls off the right hand side of the graph from the maximum of $S$ in figure 8, dynamically driven to the critical point $K_0$ at which a new domain of ecosystem relationship becomes established. This change, at an inflection point of $H$, we claim, represents a behavioral phase transition analog.

Note that, according to the analysis, transit from the right hand side of figure 8 to the left hand side, i.e. back to the original ecological configuration or domain, also faces a barrier, suggesting an origin for the quasi-stability of the deteriorated ecology.

On the other hand, that barrier may be partly surmounted from right to left, for example by policies of rebuilding devastated minority communities or of reindustrialization, until, at the peak of $S$, quite suddenly the 'richer' ecological regime again 'locks in' to quasi-stability. Thus ecosystem transitions, once they actually occur, are subject to a kind of homeostatic regulation, in this model of ecological resilience.

Note that equation (11) is expressed in terms of the effect of an embedding context – $Z$ – on a set of interacting information sources. This suggests that alteration of that context, i.e. 'resilience' transitions of the embedding 'natural' ecosystem, can entrain many of the cognitive phenomena which particularly characterize human life, both individually and in community, to put the matter euphemistically indeed.

### Discussion and conclusions

It is our contention that the many linked cognitive modules of human biology, psychology, and sociology, can become entrained by the disruption of embedding human habitat, resulting in establishment of a new quasi-stable condition producing a broad spectrum of comorbid mental and physical disorders acting at and across various levels of scale and of organization. Indeed, considerable cormorbidity should be the norm under such circumstances.

Thus this work extends ecosystem resilience theory to the cognitive phenomena which so mark humans both individually and in community, and develops, in addition, a more general tool for examining the effects of habitat disruption on the cognitive physiological and psychological submodules common across many living things. The analysis invokes an elaborate mathematical model for this, and it is crucial to make explicit the limitations of such an approach: as it is said, "all models are wrong, but some models are useful". The mathematical ecologist E.C. Pielou (1977), following the 'theoretical ecology' fiasco of the early 1970's, warns that

> "...[Mathematical] models are easy to devise; even though the assumptions of which they are constructed may be hard to justify, the magic phrase 'let us assume that...' overrides objections temporarily. One is then confronted with a much harder task: How is such a model to be tested? The correspondence between a model's predictions and observed events is sometimes gratifyingly close but this cannot be taken to imply the model's simplifying assumptions are reasonable in the sense that neglected complications are indeed negligible in their effects...
>
> In my opinion the usefulness of models is great... [however] it consists *not in answering questions but in raising them*. Models can be used to inspire new field investigations and these are the only source of new knowledge as opposed to new speculation."

In this spirit it is particularly appropriate to speculate further on the relations between public health and public policy in the homeland of the American Empire.

Figure 8 summarizes our extended perspective on ecosystem resilience suggesting that progressive degradation in human habitat – the systematic dispersal of social, economic, and political capital like the processes generating figures 1 and 4 – will, after a critical point, suddenly entrain the function of important cognitive physiological, psychological, and psychosocial modules of human biology. This, in turn, creates, a punctuated, multifactorial, multilevel, multiscale, and highly comorbid, health catastrophe, exemplified by figures 3, 6, and 7, which, in the manner of ecosystem resilience transitions, then becomes a quasi-stable, highly persistent pattern.

It is not stretching matters to call such a transition 'social eutrophication'.

While the tuberculosis outbreak of figure 3 was beaten back with a labor-intensive and expensive (about $ 40,000 per case) program of 'directly observed therapy' using highly hepatotoxic drugs, the ecological niche opened by the processes of figure 1 remains, and is likely to be filled by other diseases, both infectious and chronic. Since New York City is the strongly dominant peak of the US urban hierarchy, this circumstance has very serious implications for the national spread of such emerging infections as multiple-drug resistant

(MDR) strains of HIV and tuberculosis, as well as other possible new pathogens. Recent mayoral policies of closing more fire companies, the 'immune cells' against the contagious urban decay producing figure 1, do not bode well in this regard.

At the national scale the relentless increase of obesity and its outfalls – figure 6 – appears consequent on a persisting population-level threat from social eutrophication resulting, at the individual level, in chronic HPA axis activation and self-medication of the cortisol/leptin cycle through overeating (e.g. Bjorntorp, 2001). That pattern is, indeed, likely to worsen. The asthma epidemic among African-American communities indexed by figure 7 is unlikely to respond to medically-focused interventions which seem deliberately to ignore underlying fundamental socioeconomic and political causes driving that epidemic.

Public policies triggered the social and physical desertification of figures 1 and 4. Presumably, given the political will and a very considerable reinvestment of resources in civilian enterprise, reformist public policies can, after overcoming an inherent hysteresis, restore the human ecosystem of the US homeland to some measure of health. The alternative would appear to be further decline, sharply punctuated by episodes of comorbid social eutrophication entraining all sectors of the population. The aftermath of the collapse of the Soviet Union, which has increasingly affected even the former nomenklatura, was not an aberration, rather, according to our model, it was recognizably analogous to the events we have reviewed for the USA.

Currently predominant neoliberal ideology in the USA, however, seems to preclude intervention. The influential periodical *The Economist* (1978, p. 10), commenting on the collapse of the South Bronx, put the matter as follows:

> "The bleak truth is that this [destruction] is the *natural* and *inevitable* consequence of a shrinking city. The destruction, poverty and hopelessness that cluster around the burnt-out wrecks is abhorrent. That something should be done to stop it is the immediate reaction. That something should be done to speed it up is nearer the mark."

Norton and Rees (1979), describing the collapse of the US industrial base, express the usual view:

> "...[T]echnological changes are behind... the demise of the Manufacturing Belt. The causal links postulated between structural and regional changes are reminiscent of Schumpeter's 'process of creative destruction'..."

which they likewise see as inevitable, and indeed another 'natural' phenomenon, a position trenchantly dissected by Roberts (1991).

With such a captain and such a crew, as Melville put it in Moby Dick, even first class passengers on the American Empire seem in for a rough Middle Passage.

In spite of our elaborate biopsychosocial modeling exercise, a critically (or, less benignly, an ideologically) motivated reader might, at this point, question the relations between figures 2 and 3, or 5 and 6 and 7: are these just unrelated phenomena which happened to increase simultaneously, producing a spurious correlation? Is there really a causal relationship between deurbanization, deindustrialization, and public health decline? Administrative data exist at municipal, regional, and national levels of organization which could illuminate this question. Within New York, and probably other cities as well, data on unemployment/manufacturing employment, housing loss, and on both asthma and diabetes deaths would permit detailed spatiotemporal analysis, probably best done in five-year condensations about the 1980, 1990, and 2000 Censuses. Our earlier work on metropolitan regions (Wallace, Wallace, and Andrews, 1997) suggests that these can differ markedly in causal mechanism as well as in spatiotemporal structure and dynamic and their relations to disease – often depending on the degree of 'hollowing out' of the city center – so that analysis of health, employment, and housing data at this level of organization would provide much insight. Similarly, spatiotemporal analysis of employment, deindustrialization, population shift, and health statistics across some 3000 counties nationally would likely be useful in further exploration of national patterns, particularly lagged spatiotemporal correlations. It should prove possible to link the analyses across levels of organization, given the remarkable degree to which large cities dominate the national urban hierarchy, and central cities the urban/suburban interaction.

We call for resources to be made available for a full-scale empirical study of the long-term effects of deurbanization and deindustrialization on public health in the USA.

## Acknowledgments

This work benefited from support under NIEHS Grant I-P50-ES09600-05. The authors thank Drs. R.G. Wallace and J. Ullmann for useful discussions.

**Figure Captions**

**Figure 1**. Change in percent occupied housing units in the Bronx section of New York City between 1970 and 1980. Contagious urban decay triggered by a policy of 'planned shrinkage' involving withdrawal of fire-related municipal services created degrees of devastation in poor minority neighborhoods of the city which are unprecedented for a modern industrialized nation not at war.

**Figure 2**. Annual composite index of building fire and seriousness in New York City, 1959-1990. It is constructed by principal component analysis of number of building fires, number of 'serious' fires requiring five or more units working, and number of 'extra alarm assignments' beyond the first five units which respond. The index has been shifted so that zero fires gives a zero value, and is a measure of the fire and housing abandonment epidemic producing figure 1. This outbreak began at the 1970 level of housing overcrowding, about 70,000 housing units having more than 1.51 persons per room. By 1990, progressive housing loss had produced about 140,000 such badly overcrowded units, rewinding the ecological machinery for another possible outbreak of contagious urban decay. As of this writing – and in the aftermath of the 9/11 attack – the City has begun closing more fire companies.

**Figure 3**. Number of reported tuberculosis cases in New York City, 1959-1990, as a function of the integral of composite damage of figure 2. An integral is taken since, once a community is destroyed by contagious urban decay, its social, political, and economic capital are effectively dispersed, on the time scale of our analysis. Two 'ecological domains' are evident: decline before, and rise after, 1978. The outbreak was contained by instituting an expensive program of 'directly observed therapy', but the underlying urban ecosystem appears grossly unstable.

**Figure 4**. Counties of the Northeastern US which lost more than 1000 manufacturing jobs between 1972 and 1987. This is the infamous 'rust belt' which, in its dispersal of social, economic, and political capital, closely mirrors the devastation of figure 1.

**Figure 5**. Number of manufacturing jobs in the US between 1980 and 2001. Note the sudden decline after 1980, whose effects are well described by Pappas (1989).

**Figure 6**. US national rate of diabetes deaths per 100,000 population as a function of the *integrated* loss of manufacturing jobs after 1980. Diabetes is a good index of population obesity. As in figure 3, two ecological domains are evident: stability through 1987, followed by a sudden jump and a roughly linear rise thereafter.

**Figure 7**. African-American asthma deaths per million as a function of the same integrated manufacturing job loss as figure 6. The ecosystem transition is more spread out, showing an 'inverted-J' pattern of transition to a higher level. The health status of African-American communities may be a highly sensitive index of human habitat disruption for reasons of both concentrated poverty and the cumulative, persistent, effects of historical patterns of abuse.

**Figure 8**. The quasi-thermodynamics of resilience. As $K$ increases, the 'richness' of the associated ecology, as measured by the source uncertainty $H$, progressively declines. The 'disorder' variate $S = H - KdH/dK$ repels social and biological systems, in this model, providing a limited version of homeostasis, until essential parameters of habitat integrity are broached, and a new ecological regime becomes established as a 'phase transition' at the critical value of $K = K_0$. Note that $S$ presents a barrier both to eutrophication and to recovery from it.

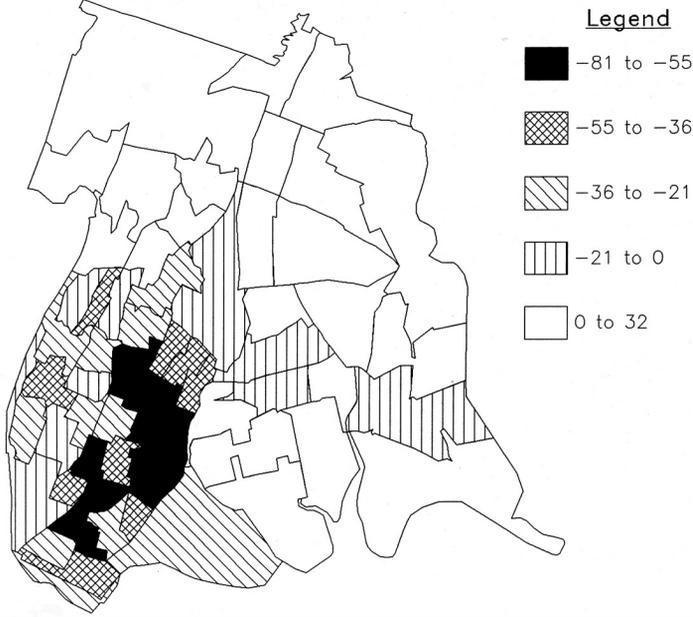

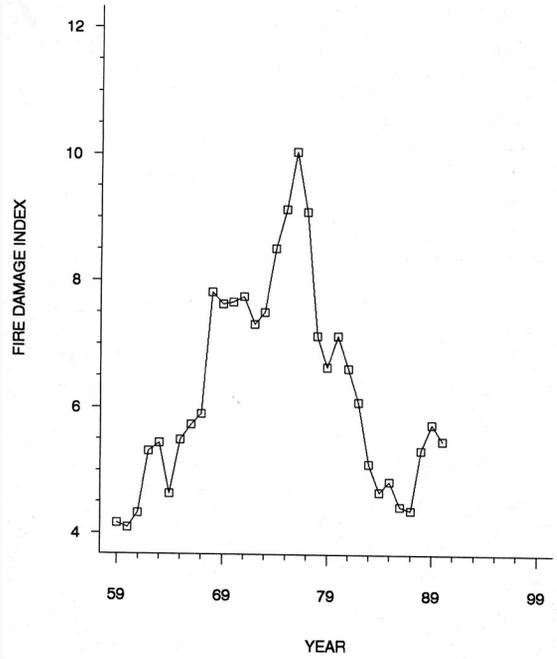

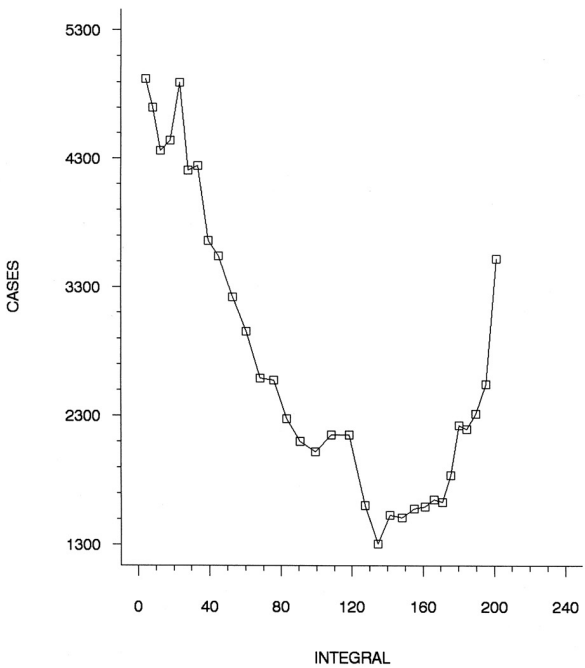

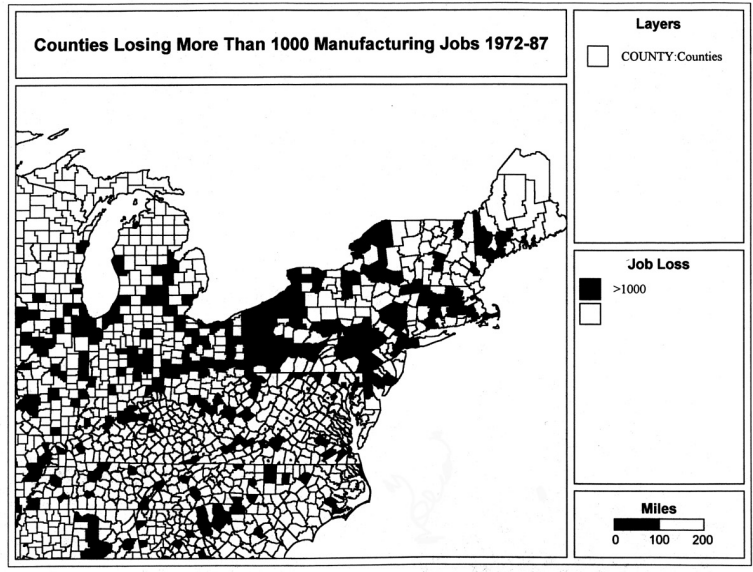

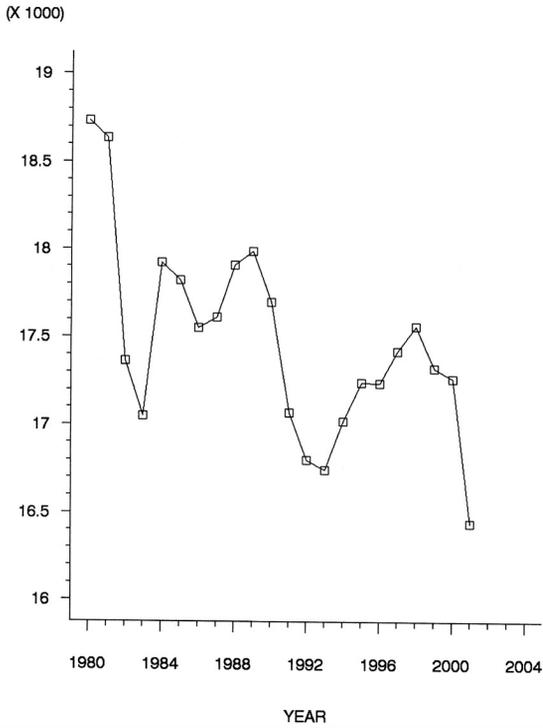

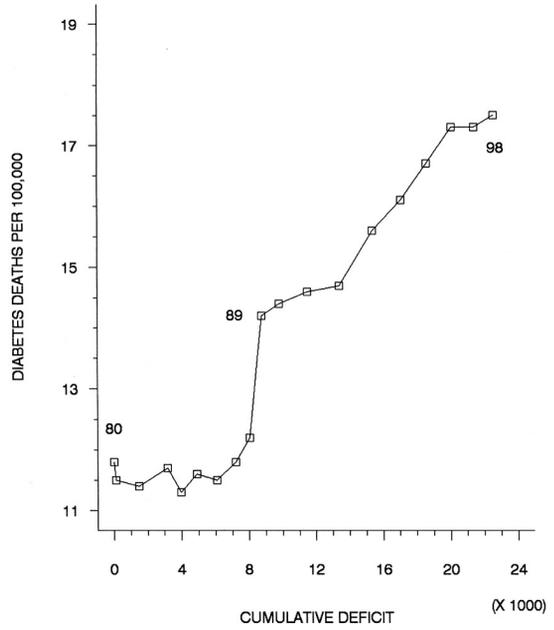

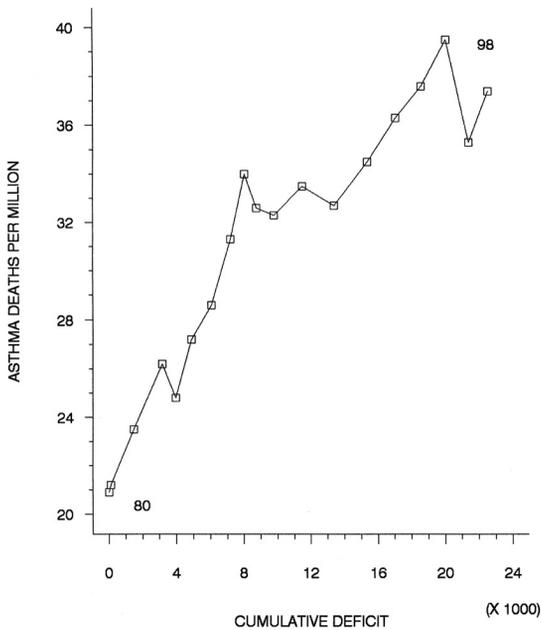

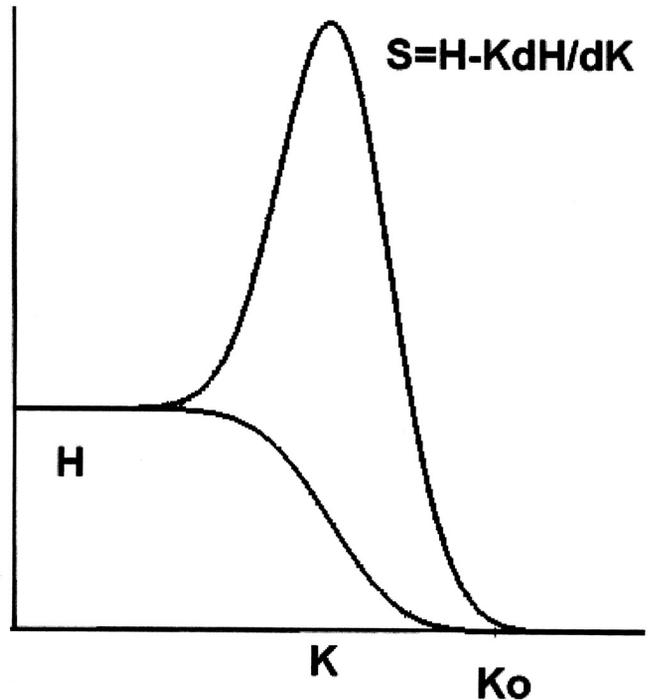